\documentclass[lettersize,journal]{IEEEtran}
\usepackage{amsmath,amsfonts}
\usepackage{algorithmic}
\usepackage{algorithm}
\usepackage{array}
\usepackage[caption=false,font=normalsize,labelfont=sf,textfont=sf]{subfig}
\usepackage{textcomp}
\usepackage{stfloats}
\usepackage{url}
\usepackage{verbatim}
\usepackage{graphicx}
\usepackage{cite}
\usepackage{bm}
\usepackage{array}
\usepackage{eso-pic}
\newcolumntype{B}{r@{\,/\,}l}
\usepackage{balance}
\usepackage{amssymb}
\usepackage{enumitem}
\usepackage{booktabs}
\usepackage{multirow}
\usepackage{multicol,makecell}
\usepackage{xspace}
\usepackage{graphicx}
\usepackage{xcolor}
\usepackage{fancyhdr}
\usepackage[normalem]{ulem}
\usepackage[bottom]{footmisc}
\usepackage{courier}
\usepackage{longfbox}
\usepackage{caption}
\usepackage{xurl}                  %
\usepackage[]{hyperref}
\usepackage[colorinlistoftodos,prependcaption,textsize=scriptsize]{todonotes} %
\usepackage{makecell}

\usepackage[sort&compress]{cleveref}
\usepackage{xstring}               %
\usepackage{pifont}   %

\usepackage{xparse}      %
\usepackage{expl3}       %

\usepackage{tikz}

\usepackage{comment}

\usepackage{kotex}

\usepackage[most]{tcolorbox}

\definecolor{pred}{rgb}{0.7843, 0.0039, 0.3137} %
\definecolor{darkred}{rgb}{0.8, 0, 0}
\definecolor{darknavy}{rgb}{0, 0, 0.5}
\definecolor{darkgreen}{rgb}{0, 0.4, 0}
\definecolor{darkpurple}{RGB}{102,51,153}
\definecolor{blue}{rgb}{0, 0, 0}
\definecolor{darkorange}{rgb}{1, 0.6, 0.3}

\newcommand{\rev}[1]{{\color{blue}{#1}}}

\newcommand{\HBMfour}{\textsf{HBM4}\xspace}

\newcommand{\HBFcons}{\textsf{CONV}\xspace}
\newcommand{\HBFconsplus}{\textsf{CONV}$^\textsf{+}$\xspace}

\newcommand{\HBF}{\textsf{HBF}\xspace}
\newcommand{\HBFplus}{\textsf{HBF}$^\textsf{+}$\xspace}

\newcommand{\bw}[2]{%
  \makebox[2.8em][r]{#1}\,/\,\makebox[2.8em][l]{#2}%
}

\newcounter{take}
\newcommand{\take}[1]{%
  \par\vspace{-5pt}%
  \refstepcounter{take}%
  \begin{tcolorbox}[enhanced,
    colback=lightgray!35!white,   %
    boxrule=0.5pt,                  %
    arc=4pt,                      %
    left=0pt,right=0pt,top=0.1pt,bottom=0pt %
  ]%
    \head{Takeaway~\thetake}\space{#1}%
  \end{tcolorbox}%
  \par\vspace{-5pt}%
  \noindent\ignorespaces
}

\newcommand{\densefull}{\text{Llama 3 405B}\xspace}
\newcommand{\moefull}{\text{Llama 4 Maverick}\xspace}
\newcommand{\dense}{\text{Llama3}\xspace}
\newcommand{\moe}{\text{Llama4}\xspace}

\newcommand{\wshort}{\textsf{\textsc{Short}\xspace}}
\newcommand{\wmid}{\textsf{\textsc{Mid}\xspace}}
\newcommand{\wlong}{\textsf{\textsc{Long}\xspace}}

\newcommand{\inlength}{$L_\text{IN}$\xspace}
\newcommand{\outlength}{$L_\text{OUT}$\xspace}

\newcommand{\fig}[1]{{Fig.~#1}\xspace}

\newcommand{\tbl}[1]{{Table~#1}\xspace}

\newcommand\inum[1]{(\textit{#1})\xspace}
\newcommand{\head}[1]{{\noindent\textbf{#1.}\xspace}}

\newcommand{\usec}{\textmu{}s\xspace}

\newcommand{\IEEEArxivNotice}{%
\AddToShipoutPictureFG*{%
  \AtPageUpperLeft{%
    \raisebox{-12pt}{%
      \makebox[\paperwidth][c]{%
        \parbox{0.94\paperwidth}{\centering\scriptsize
        This article has been accepted for publication in IEEE Computer Architecture Letters.
        This is the author's version which has not been fully edited and content may change prior to final publication.
        Citation information: DOI 10.1109/LCA.2026.3705817.
        }%
      }%
    }%
  }%
  \AtPageLowerLeft{%
    \raisebox{12pt}{%
      \makebox[\paperwidth][c]{%
        \parbox{0.94\paperwidth}{\centering\scriptsize
        \copyright~2026 IEEE. Personal use of this material is permitted.
        Permission from IEEE must be obtained for all other uses, in any current or future media,
        including reprinting/republishing this material for advertising or promotional purposes,
        creating new collective works, for resale or redistribution to servers or lists,
        or reuse of any copyrighted component of this work in other works.
        }%
      }%
    }%
  }%
}%
}

\begin{document}

\IEEEArxivNotice

\title{\huge Exploring High-Bandwidth Flash for Modern LLM Inference: Opportunities and Challenges}

\author{\resizebox{\linewidth}{!}{\small Dowon Son, Yonggon Park, Hyunuk Cho, Hyungkyu Ham, Onur Mutlu, Sungjin Lee, Gwangsun Kim, and Jisung Park}
\vspace{-1em}
\thanks{D. Son, Y. Park, H. Cho, H. Ham, S. Lee, G. Kim, and J. Park are with POSTECH, Pohang-si, Gyeongsangbuk-do, Republic of Korea.}
\thanks{O. Mutlu is with ETH Z\"urich, Zurich, Switzerland.}
}

\setlength{\skip\footins}{2pt plus 1pt minus 1pt}
\setlength{\intextsep}{2pt plus 1pt minus 1pt}
\setlength{\textfloatsep}{2pt plus 1pt minus 1pt}

\IEEEaftertitletext{\vspace{-2.5\baselineskip}}
\maketitle

\begin{abstract}

This work investigates the potential benefits and technical challenges of using high-bandwidth flash (HBF) for large language model (LLM) inference.
HBF has gained increasing attention as a promising solution to mitigate memory-capacity bottlenecks in modern LLM-serving systems, but its benefits and challenges remain largely uninvestigated.
To address this gap, we thoroughly analyze HBF-based LLM-serving systems under diverse system configurations and operating scenarios in which HBF serves as a main GPU-memory component to handle both reads and writes.
Our analysis shows that, despite its limited write performance, HBF can significantly improve the batch size, throughput, and flexibility of LLM-serving systems while reducing the minimum GPU requirements, but realizing these benefits critically depends on sustaining HBM-comparable read bandwidth and requires significant endurance improvements.

\end{abstract}
\begin{IEEEkeywords}
High-bandwidth flash, LLM-serving systems
\end{IEEEkeywords}

\section{Introduction} \label{sec:intro} \thispagestyle{empty}

\IEEEPARstart{M}{odern} computing systems for LLM inference suffer from the limited capacity of HBM.
To improve inference quality, recent LLMs continue to grow rapidly in model size and context length, far outpacing HBM-capacity scaling.
A common solution to this mismatch is to scale aggregate HBM capacity with multiple GPUs, which \inum{i}~significantly
increases system cost by requiring many GPUs and expensive high-speed interconnects and \inum{ii}~limits system flexibility by tightly coupling multiple GPUs to serve each query.

HBF has gained increasing attention as a promising solution to mitigate the memory-capacity bottleneck.
It replaces HBM's DRAM with low-latency NAND flash dies, significantly increasing per-stack capacity by more than 14$\times$ (e.g., 512~GB) compared to HBM4 (36~GB).
By exploiting massive die- and plane-level parallelism, HBF aims to deliver per-stack read bandwidth comparable to HBM4 (e.g., 1.6~TB/s~\cite{ma-arXiv-2026}).

In this work, we present the first systematic study on the opportunities and challenges of using HBF to \emph{replace} (most of) HBM in modern LLM-serving systems.
Despite the significant capacity advantage over HBM, HBF's effectiveness as a main GPU-memory component in practical LLM-serving scenarios remains largely unexplored.
To address this, we thoroughly analyze HBF-based LLM serving with recent large-scale models while varying GPU counts, target service-level objectives (SLOs), and average context lengths.

Based on our analysis, we make ten key observations and five takeaways on the performance, bottlenecks, and design implications of HBF-based LLM-serving systems.
We highlight three key findings.
First, HBF has strong potential to significantly improve the batch size, throughput, and flexibility of modern LLM-serving systems, despite its limited write performance.
Second, the larger memory capacity alone is insufficient to realize these benefits; sustaining HBM-comparable read bandwidth is also critical under SLO constraints.
Third, in HBF-based systems, KV-cache writes introduce non-trivial performance overheads and significant endurance challenges.

\section{Background and Motivation} \label{sec:background}

\begin{figure}[h]
     \centering
     \includegraphics[width=0.95\linewidth]{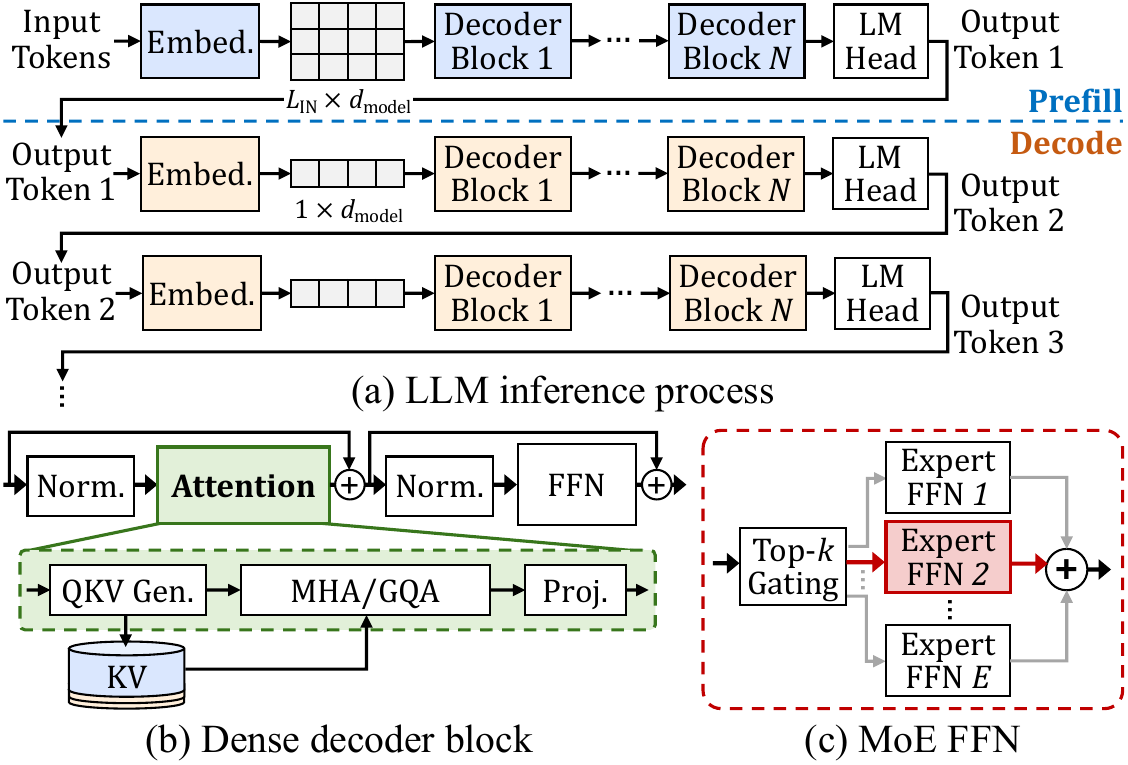}
     \caption{LLM inference process and architecture.}
     \label{fig:llm_architecture}
\end{figure}

\head{LLM Inference}
\fig{\ref{fig:llm_architecture}(a)} depicts the overall process of LLM inference, which consists of two phases: \inum{i}~\emph{prefill} and \inum{ii}~\emph{decode}.
Both phases pass through the same token-embedding layer, decoder blocks, and language-model (LM) head.
Given a query with \inlength tokens, the prefill phase processes all tokens in parallel to generate the first output token.
The decode phase then generates \outlength tokens \emph{auto-regressively}, using each output token as a new input token for the next iteration.

The decoder blocks iteratively refine each token's representation, i.e., a $d_\text{model}$-dimensional hidden vector, whereas the token-embedding layer and LM head map between tokens and vectors.
Each decoder block contains two key components: \inum{i}~an \emph{attention layer} and \inum{ii}~a \emph{feedforward network} (FFN) layer (\fig{\ref{fig:llm_architecture}(b)}).
The attention layer projects input representations into query, key, and value (QKV) tensors and aggregates information via multi-head attention (MHA);
it is common practice to cache all key and value tensors (KV cache) to avoid redundant recomputation in the decode phase.
The FFN layer applies position-wise nonlinear transformations independently to each token.
In dense models (e.g., \densefull~\cite{grattafiori-arXiv-2024}), a single large FFN processes all tokens, whereas recent mixture-of-experts (MoE) models (e.g., \moefull~\cite{llama4}) use multiple smaller expert FFNs and route each token to only a few selected ones, enabling large total model capacity while limiting per-token computation and memory access (\fig{\ref{fig:llm_architecture}(c)}).

\head{High-Bandwidth Flash}
Since its introduction in early 2025~\cite{HBF_release}, HBF has attracted growing attention as a potential solution to the memory-capacity bottleneck in modern LLM-serving systems.
Recent LLMs have rapidly grown far beyond a single GPU's memory capacity (e.g., \moefull{}'s model size is 746~GB, whereas a high-end GPU~\cite{Rubin} provides only 288~GB), making multi-GPU deployment inevitable.
Although various optimizations such as quantization~\cite{hooper-arXiv-2025} and grouped-query attention (GQA)~\cite{ainslie-EMNLP-2023} significantly reduce KV-cache size, the maximum batch size remains limited due to continuously increasing model size and context length.
HBF has the potential to fundamentally address this bottleneck by exploiting the high storage density of NAND flash memory.

\begin{figure}[t]
     \centering
     \includegraphics[width=.9\linewidth]{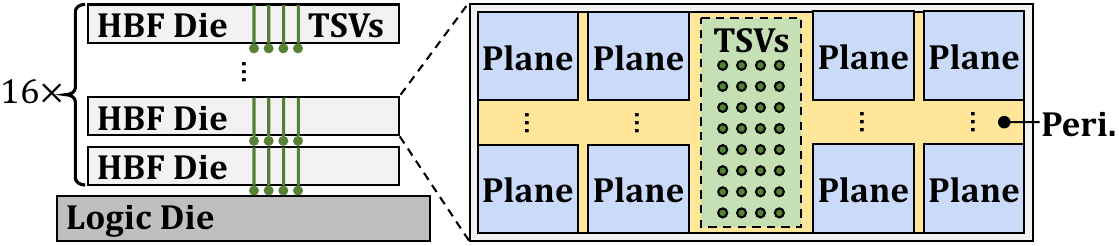}
     \caption{Expected organization of HBF.}
     \label{fig:HBF_organization}
\end{figure}

\fig{\ref{fig:HBF_organization}} shows the high-level organization of an HBF stack, which vertically stacks multiple core dies on a logic base die via TSVs, as in conventional HBM.
Each core die contains multiple low-latency NAND flash planes that operate concurrently, thereby providing high read bandwidth through massive parallelism.
According to Sandisk’s projection, HBF targets \usec-order latency for 4-KiB reads while delivering per-stack bandwidth comparable to that of HBM4 (e.g., 1.6~TB/s)~\cite{ma-arXiv-2026}.
\rev{An HBF stack is expected to consume higher power ($<$ 80~W) compared to an HBM stack (40~W)~\cite{ma-arXiv-2026}, which could also cause thermal issues.
However, we believe that this is primarily a power-delivery and thermal-management challenge rather than a fundamental limitation of HBF.
Note that, as we demonstrate in this work, HBF has the potential to reduce the GPU count required for serving large-scale LLMs, which may help offset its higher per-stack power consumption at the system level.
}

\textbf{Our goal} in this work is to provide new insights into HBF's practical benefits and technical challenges in modern LLM-serving systems, which remain largely underexplored despite HBF's high potential.
To our knowledge, only one recent work has shown HBF's effectiveness at improving the efficiency of cache-augmented generation (CAG) when used as a read-only memory extension to HBM~\cite{ha-CAL-2026}, but it is unclear how HBF can benefit more general LLM-serving scenarios as a main GPU-memory component. To address this, we thoroughly analyze the effectiveness of HBF-enabled LLM-serving systems with various system configurations and workloads.

\section{Methodology} \label{sec:methodology}

\head{Hardware Configuration}
\tbl{\ref{tab:gpu_configs}} summarizes the GPU-memory configurations of five LLM-serving systems we evaluate, which share the same state-of-the-art GPU-core architecture~\cite{Rubin}.
First, \HBMfour contains eight HBM stacks per GPU, each providing 36-GB capacity and 1.6-TB/s read bandwidth.

Second, \HBF and \HBFplus replace seven and all eight HBM stacks of the GPUs in \HBMfour, respectively, with HBF stacks offering 512~GB of capacity and 1.6~TB/s of read bandwidth per stack as projected by Sandisk; 
to avoid excessive writes to NAND flash memory, we assume that \HBF reserves one HBM stack to store intermediate data, whereas \HBFplus leverages 40-MB SRAM on the logic die instead.\footnote{We find that writing intermediate data increases write traffic by up to 100$\times$ for every decode iteration, incurring significant endurance and performance overheads. We use the SRAM capacity from prior work~\cite{ha-CAL-2026}.}
\rev{Our analysis shows that much of the intermediate data has a short lifetime, which enables quick release of SRAM-buffer space and thus allows \HBFplus to support a large batch size (e.g., 327 under \moefull) despite the limited SRAM capacity (320 MB per GPU).
Both \HBF and \HBFplus place model weights and KV cache in NAND flash, as they require a large storage capacity and exhibit read-dominant access patterns.
To achieve the target read bandwidth (1.6~TB/s per stack), we assume 25 planes per HBF core die and a 1-\usec 4-KiB page-read latency.
As in prior work~\cite{ha-CAL-2026}, we further assume that each HBF stack uses a dedicated 3.13-MB SRAM staging buffer on its logic die to prefetch and double-buffer read data, thereby minimizing the performance impact of \usec-scale read latency.}

Third, \HBFcons and \HBFconsplus are identical to \HBF and \HBFplus, respectively, except for each stack's read and write bandwidth.
We introduce these conservative configurations \rev{since achieving HBF's target bandwidth requires significant advancements over existing low-latency NAND flash technologies, e.g., Samsung's Z-NAND~\cite{cheong-ISSCC-2018} and Kioxia's XL-Flash~\cite{kouchi-ISSCC-2020} that provide only 87.4-GB/s and 262-GB/s read bandwidth with 16 dies, respectively.
Although future HBF could achieve the target bandwidth through \inum{i}~finer-grained plane partitioning, which would reduce read latency and increase internal parallelism, and \inum{ii}~wafer-bonding-based integration~\cite{HBF_spec}, which would mitigate the area overhead of aggressive partitioning, we project \HBFcons and \HBFconsplus from existing low-latency NAND flash technologies.}
We assume 16 planes per die~\cite{kouchi-ISSCC-2020} and a 3-\usec page-read latency~\cite{cheong-ISSCC-2018}, achieving per-stack read bandwidth of 0.35~TB/s. 
Since HBF's expected write performance is not publicly available, we assume a 100-\usec page-program latency for all HBF configurations based on~\cite{cheong-ISSCC-2018}.

\begin{table}[t]
  \caption{GPU-memory configurations of evaluated systems.}
  \label{tab:gpu_configs}
  \centering
  \setlength{\tabcolsep}{4pt}
  \resizebox{\columnwidth}{!}{%
    \begin{tabular}{ccccc}
      \toprule
      \textbf{GPU} &
      \textbf{\makecell{\# Stacks \\ (HBM/HBF)}} &
      \textbf{\makecell{Capacity {[GB]}}} &
      \textbf{\makecell{RD BW {[TB/s]} \\ (HBM/HBF)}} &
      \textbf{\makecell{WR BW {[TB/s]} \\ (HBM/HBF)}} \\
      \midrule
      \HBMfour     & 8/0 &   288 & \bw{12.8}{-}   & \bw{12.8}{-}    \\
      \HBFcons     & 1/7 & 3,620 & \bw{1.6}{2.45} & \bw{1.6}{0.073} \\
      \HBFconsplus & 0/8 & 4,096 & \bw{-}{2.80}   & \bw{-}{0.084}   \\
      \HBF         & 1/7 & 3,620 & \bw{1.6}{11.2} & \bw{1.6}{0.112} \\
      \HBFplus     & 0/8 & 4,096 & \bw{-}{12.8}   & \bw{-}{0.128}   \\
      \bottomrule
    \end{tabular}%
  }
\end{table}

\head{Simulation Methodology}
We use an existing simulator for LLM-serving systems, called LLMSimulator~\cite{yun-MICRO-2024}, which we extend in three ways.
First, we add an HBF-aware timing model that individually configures read and write bandwidths to reflect the performance asymmetry of NAND flash memory.
Second, we improve the modeling of continuous batching and disaggregated prefill-decode execution by accounting for the performance overhead of KV-cache writes when transferring outputs from prefill to decode nodes.
Third, we implement key techniques widely adopted in modern LLMs and LLM-serving systems, including chunked attention and pipeline parallelism.
We will open-source our extended LLM simulator.

We make four assumptions about the system configuration and execution model in our analysis.
First, we focus on decode nodes in disaggregated prefill-decode execution, as the prefill phase is inherently compute-bound and thus benefits little from HBF’s large capacity.
Second, we assume continuous batching that maintains a full batch by admitting new queries as in-flight ones complete.
Third, each evaluated system selects the parallelism configuration that maximizes the achievable system throughput subject to all constraints, including SLO requirements (e.g., time per output token (TPOT) $\leq$ 0.1 second), GPU-memory capacity, and on-die SRAM limits (in \HBFplus and \HBFconsplus), by combining data, tensor, pipeline, and expert parallelism.
Fourth, all evaluated systems are equipped with high-performance interconnects: NVLink (1,800~GB/s) for intra-node GPU communication (up to eight GPUs per node) and InfiniBand (100 GB/s) for inter-node communication.

\head{Workloads}
We evaluate \densefull (dense) and Llama~4 Maverick (MoE) across three application scenarios with different context-length characteristics;
we vary each query's $\langle{L_\text{IN}}$, $L_\text{OUT}\rangle$ values across $\langle$1,660, 373$\rangle$, $\langle$5.9K, 499$\rangle$, and $\langle$103.5K, 1.1K$\rangle$ based on the average values in the shareGPT~\cite{sharegpt-gpt4} (\wshort), LongBench~\cite{bai-arXiv-2024} (\wmid), and English summarization~\cite{zhang-arXiv-2024} (\wlong), respectively.
We model a constant query arrival rate matching the completion rate, which enables steady-state performance analysis under continuous batching.

\section{Evaluation Results} \label{sec:results}

\begin{figure}[h]
     \centering\includegraphics[width=\linewidth]{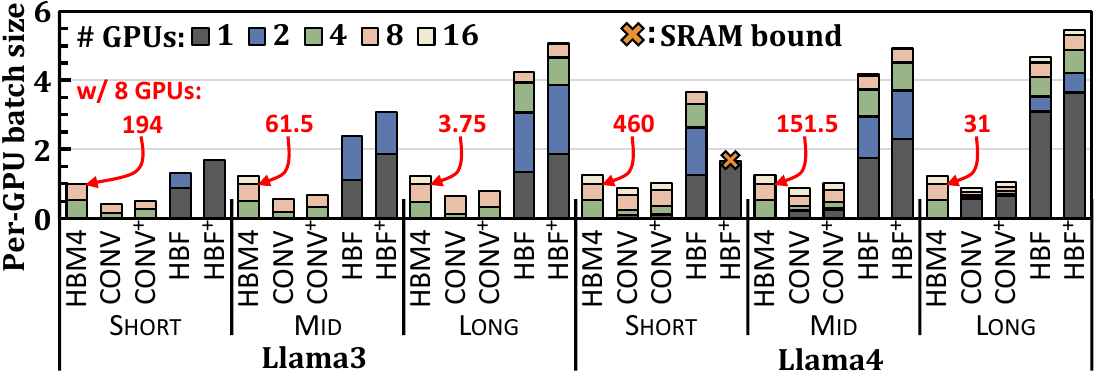}
     \caption{Per-GPU batch size under 0.1-second TPOT SLO.} 
     \label{fig:batch_size}
\end{figure}

\head{Batch Size} 
\fig{\ref{fig:batch_size}} compares the maximum \emph{per-GPU} batch sizes of five LLM-serving systems under 0.1-second TPOT SLO across GPU counts ranging from 1 to 16.
We compactly encode results for different GPU counts within a single bar;
the top of each colored segment indicates the maximum per-GPU batch size at the corresponding GPU count.
\rev{If a larger GPU count does not increase per-GPU batch size, the corresponding segment does not appear in \fig{\ref{fig:batch_size}} because it is subsumed by the preceding segment.}
All values are normalized to 8-GPU \HBMfour for each workload, whose values are denoted in~\fig{\ref{fig:batch_size}}.

We make four key observations from \fig{\ref{fig:batch_size}}.
First, \HBF and \HBFplus enable significantly larger batch sizes than \HBMfour at the same GPU count while meeting the TPOT SLO across all workloads, e.g., providing
1.3--4.5$\times$ and 1.7--5.3$\times$ gains, respectively, in the 8-GPU configuration.
In particular, \HBMfour requires at least four GPUs to serve both LLMs (\emph{no} 1-/2-GPU segments in \emph{all} \HBMfour bars), whereas 1-GPU \HBF and \HBFplus support a larger per-GPU batch size than 8-GPU \HBMfour in most cases.
The results suggest that HBF has high potential to enable more cost-effective LLM-serving systems by substantially reducing the minimum GPU requirement.

Second, \HBFcons and \HBFconsplus support only limited batch sizes, often even smaller than those of \HBMfour.
Their relatively low read bandwidths make it challenging to meet the TPOT SLO with large batches, leading to significant underutilization of the HBF capacity (e.g., up to 95\% with \dense under \wshort).
The results demonstrate that increasing memory capacity alone is insufficient for practical HBF-based LLM-serving systems; sustaining read bandwidth comparable to HBM is critical to take advantage of the expanded capacity.

Third, dedicating an HBM stack to intermediate data limits the batch-size improvements of HBF-based systems, e.g., with eight GPUs, \HBFplus can support 24\% larger batch sizes on average than \HBF in \wmid.
In \HBF, the single HBM stack stores only a limited amount of data, so its bandwidth remains largely underutilized while the HBF stacks handle most data accesses. 
These results again highlight the importance of read bandwidth to fully exploit HBF under SLO constraints.

Fourth, \HBF and \HBFplus provide increasing benefits with the context length, e.g., \HBFplus achieves 124\%/200\% higher average gains over \HBMfour in \wmid/\wlong{} than in \wshort.
In general, LLM-serving systems can support larger batches for shorter-context queries, which limits HBF's benefits in three aspects;
\inum{i}~the inter-GPU communication increases almost linearly with batch size, making it difficult to further scale the already large batch sizes in short-context workloads under the TPOT SLO;
\inum{ii}~in the dense model, FFN execution shifts toward compute-bound GEMM (general matrix multiplication) operations, which also limits further batch-size scaling under the SLO; 
\inum{iii}~a larger batch size also linearly increases the peak size of intermediate data, so the limited SRAM-buffer capacity (e.g., 40 MB) bottlenecks batch-size scaling in \HBFplus.

\take{HBF has the potential to significantly increase batch sizes in modern LLM inference, but its practical benefits over HBM depend critically on its read bandwidth.}

\begin{figure}[t]
     \centering\includegraphics[width=\linewidth]{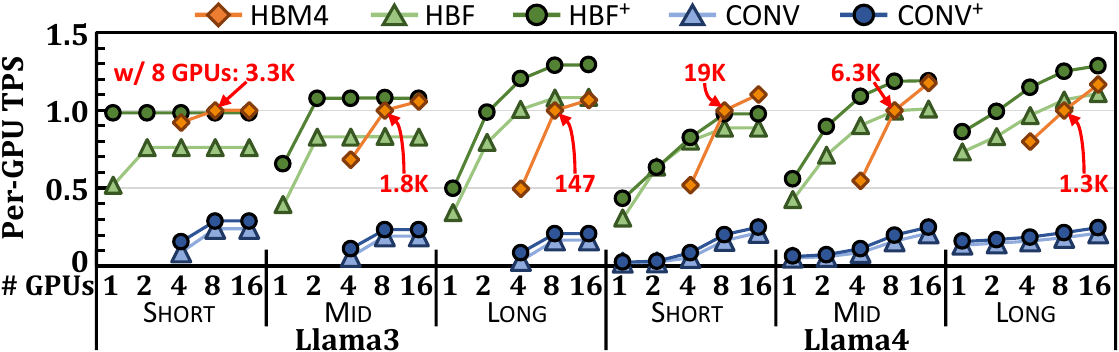}
     \caption{System throughput of evaluated LLM-serving systems.}
     \label{fig:system_throughput}
\end{figure}

\head{System Throughput} 
\fig{\ref{fig:system_throughput}} compares the per-GPU token-generation throughput, i.e., tokens per second (TPS), of five LLM-serving systems under 0.1-second TPOT SLO across GPU counts ranging from 1 to 16.
All values are normalized to the 8-GPU \HBMfour (denoted) for each workload. 
We make two key observations from \fig{\ref{fig:system_throughput}}.
First, for each evaluated LLM and workload, both \HBF and \HBFplus provide substantially higher per-GPU TPS than \HBMfour at one or more GPU counts, despite their limited write performance.
Although their gains diminish as GPU count increases, this does \emph{not} imply that HBF benefits only small-scale systems.
In fact, \HBMfour also eventually exhibits diminishing TPS gains from GPU scaling, as shown in \fig{\ref{fig:system_throughput}}; 
larger batches improve TPS by better amortizing model-weight accesses across more queries, but beyond a sufficient batch size, the benefits severely decrease as the bottleneck shifts to other operations whose costs increase with the batch size, e.g., KV-cache accesses and inter-GPU communication.
\HBF and \HBFplus consistently achieve such batch sizes with fewer GPUs and thus provide significant TPS gains until \HBMfour is scaled to sufficient GPUs, which suggests that HBF can benefit a wide range of LLM-serving scenarios, e.g., enabling large-scale systems to better cope with the continuous, rapid growth in model size.

Second, \HBFplus can provide higher per-GPU TPS than \HBMfour \emph{even with fewer GPUs}.
For example, in the \moe \wlong{} workload, the per-GPU TPS of 4-GPU \HBFplus is 15\% higher than that of 8-GPU \HBMfour, highlighting HBF's potential to improve the flexibility of LLM-serving systems; with eight GPUs, \HBFplus can run two independent 4-GPU instances with higher aggregate throughput than 8-GPU \HBMfour, while reducing scheduling complexity and improving fault isolation.

\take{Despite its limited write performance, HBF has the potential to improve the throughput and flexibility of LLM-serving systems while reducing the minimum GPU requirement, when used as a main GPU-memory component.}

\begin{figure}[t]
     \centering\includegraphics[width=\linewidth]{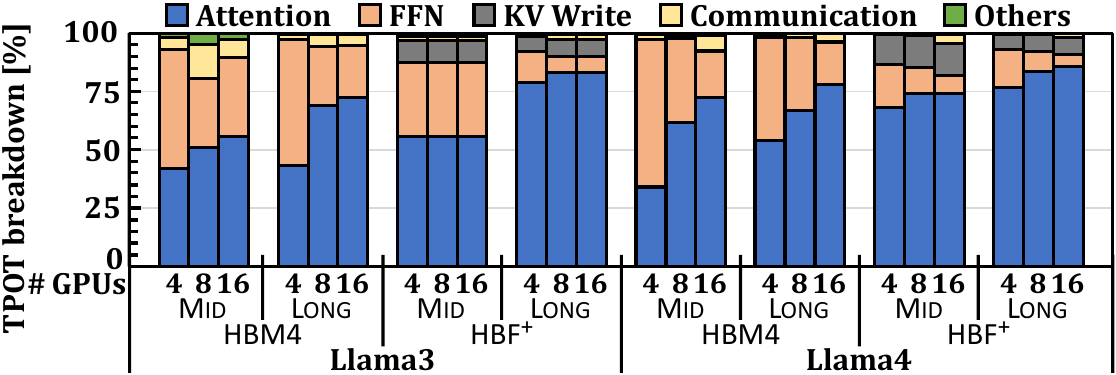}
     \caption{Runtime breakdown under 0.1-second TPOT SLO.}
     \label{fig:breakdown}
\end{figure}

\head{Performance Breakdown} 
To better understand the source of HBF's benefits over HBM, we analyze the latency breakdown of \HBFplus and \HBMfour under 0.1-second TPOT SLO.
\fig{\ref{fig:breakdown}} shows the fraction of decoding time spent on each component. `KV Write' represents writing the KV cache for \emph{newly admitted queries} under continuous batching,\footnote{Writing the KV cache newly generated during the decode phase can be overlapped with computation in the attention layer.}
and `Others' includes layer normalization, residual, and LM head.

We make two key observations from \fig{\ref{fig:breakdown}}.
First, KV-cache writes have a non-trivial performance impact in \HBFplus, e.g., accounting for 5--13.9\% of the execution time in Llama4.
The impact increases with more GPUs, shorter contexts, and the MoE model, due to more frequent query admissions.
The results suggest that \HBFplus could provide even greater benefits over \HBMfour with improved write performance.

Second, \HBFplus significantly reduces the fraction of decoding time spent on FFN layers and inter-GPU communications.
Compared to \HBMfour, \HBFplus consistently shifts the performance bottleneck toward attention layers, suggesting that it \inum{i}~more effectively amortizes model-weight accesses across queries with larger batches and \inum{ii}~reduces communication overheads by leveraging less tightly coupled inter-GPU parallelism.
In attention layers, massive KV-cache reads dominate execution time, diminishing the benefits of large batches.

\take{Improving KV-cache write performance can further increase the benefits of HBF-based systems.}

\begin{figure}[h]
     \centering\includegraphics[width=\linewidth]{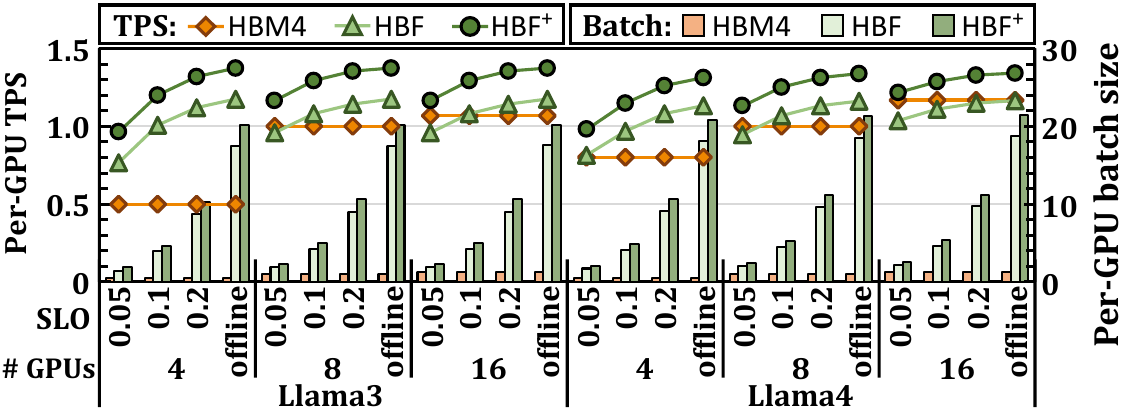}
     \caption{Performance and batch size impact of TPOT SLO.}
     \label{fig:effect_of_SLO}
\end{figure}

\head{Effect of SLO}
We evaluate how HBF's effectiveness changes depending on the TPOT SLO.
\fig{\ref{fig:effect_of_SLO}} compares the per-GPU TPS (corresponding to the left $y$-axis) and batch size (the right $y$-axis) of three LLM-serving systems, which we measure in the \wlong{} workload while varying TPOT SLOs.
We also evaluate an offline inference scenario, where each query only needs to complete within 24 hours (i.e., no TPOT SLO).
All values are normalized to the 8-GPU \HBMfour for each workload.

We observe that the target SLO significantly affects HBF's benefits; the more relaxed the SLO, the greater the benefits HBF provides.
Although \HBFplus always outperforms \HBMfour even under 0.05-second TPOT SLO in \wlong{}, its TPS gains substantially increase in the offline scenario even at large GPU counts, e.g., from 4.1\% to 14.8\% in \moe with 16 GPUs.
The results highlight the distinct performance bottlenecks in HBM- and HBF-based systems; \HBMfour's per-GPU TPS and batch size hardly change across SLOs due to the memory-capacity bottleneck, whereas \HBF and \HBFplus significantly benefit from relaxed SLOs by leveraging their large capacity.

\take{HBF is especially promising for throughput-oriented or offline LLM inference with relaxed target SLOs.}

\begin{figure}[h]
     \centering\includegraphics[width=\linewidth]{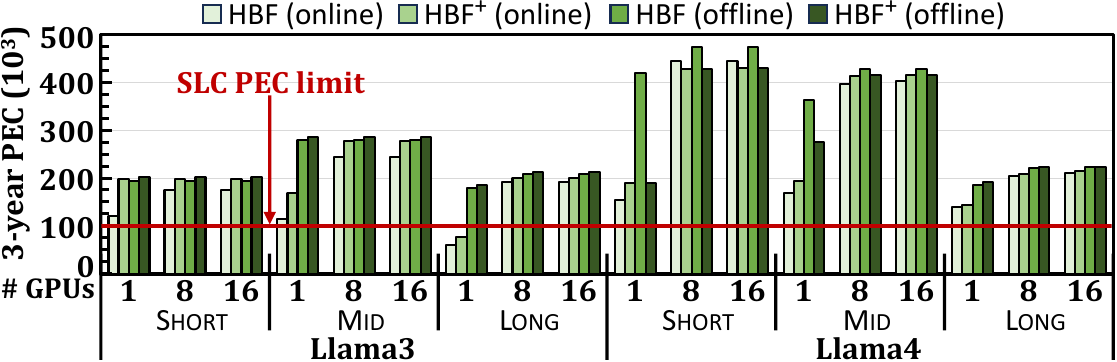}
     \caption{Required PEC under different operating scenarios.}
     \label{fig:pec}
\end{figure}

\head{Write Traffic Analysis}
We analyze the write traffic to HBF to understand the endurance requirement in HBF-based LLM-serving systems;
NAND flash memory has limited endurance, e.g., a single-level cell (SLC) block cannot guarantee data reliability after experiencing 100K program and erase (P/E) cycles.
\fig{\ref{fig:pec}} shows the per-block P/E-cycle count (PEC) of \HBF and \HBFplus over three years, a common GPU-warranty period, under various operating scenarios.

We observe that both \HBF and \HBFplus incur significant P/E-cycle counts far exceeding 100K in most cases.
In general, the write traffic increases more sharply when the systems can achieve higher TPS, i.e., with shorter context lengths, more GPUs, relaxed target SLO, and the MoE model, as every processed token generates its own KV-cache tensors.
Writing the KV cache in a retention-relaxed manner~\cite{pan-HPCA-2025} could mitigate the endurance problem, but we expect that significant improvements in NAND flash endurance are still critical to prevent the premature replacement of costly GPUs.

\take{NAND flash endurance remains a critical challenge for practical HBF-based LLM-serving systems.}

\section{Conclusion} \label{sec:conclusion}
We present the first study on the opportunities and challenges of HBF as an alternative to HBM in modern LLM-serving systems.
Our analysis shows that HBF has strong potential to improve the throughput and
flexibility while reducing the minimum GPU requirements, but it also requires HBM-comparable read bandwidth as well as significant improvements in write performance and endurance.
We hope that our findings help guide future research toward efficient and practical HBF-based LLM-serving systems.

\bibliographystyle{IEEEtran}
\bibliography{refs}

\end{document}